%% file: main.tex
\documentclass[journal]{IEEEtran}
\usepackage{amsmath,amsfonts}
\usepackage{algorithmic}
\usepackage{algorithm}
\usepackage{array}
\usepackage[caption=false,font=normalsize,labelfont=sf,textfont=sf]{subfig}
\usepackage{textcomp}
\usepackage{stfloats}
\usepackage{url}
\usepackage{verbatim}
\usepackage{graphicx}
\usepackage{cite}
\usepackage{booktabs} 
\usepackage{multirow}
\usepackage{threeparttable}
\usepackage{bm}
\usepackage{xcolor}
\usepackage{tabularx}
\usepackage{array}
\usepackage{CJKutf8}

\newtheorem{remark}{Remark}

\begin{document}
\begin{CJK}{UTF8}{gbsn}

\title{A Comprehensive Overview of Backdoor Attacks in Large Language Models within Communication Networks}

\author{Haomiao~Yang,~\IEEEmembership{Member,~IEEE,}
        Kunlan~Xiang,
        Mengyu~Ge,~\IEEEmembership{Graduate Student Member,~IEEE,}
        Hongwei~Li,‬~\IEEEmembership{Senior Member,~IEEE,}
        Rongxing~Lu,~\IEEEmembership{Fellow,~IEEE}
        Shui~Yu,~\IEEEmembership{Fellow,~IEEE}}

\maketitle


\begin{abstract}
\input{abstract}
\end{abstract}


\section{Introduction}
\input{Introduction}

\section{PRELIMINARIES}
\label{pre}
\input{preliminary}

\section{THREAT MODEL}
\label{threatModel}
\input{threat_model}

\section{Backdoor Attacks in LLMs}
\label{backdoors}
\input{backdoors}

\section{Benchmark Datasets}
\label{Benchmark Datasets}
\input{Benchmark_Datasets}

\section{FUTURE RESEARCH DIRECTIONS}
\label{future_research}
\input{future_directions}

\section{CONCLUSION}
\label{conclusion}
\input{conclusion}

\newpage
\end{CJK}
\end{document}

%% file: abstract.tex
The Large Language Models (LLMs) are poised to offer efficient and intelligent services for future mobile communication networks, owing to their exceptional capabilities in language comprehension and generation. However, the extremely high data and computational resource requirements for the performance of LLMs compel developers to resort to outsourcing training or utilizing third-party data and computing resources. These strategies may expose the model within the network to maliciously manipulated training data and processing, providing an opportunity for attackers to embed a hidden backdoor into the model, termed a backdoor attack. Backdoor attack in LLMs refers to embedding a hidden backdoor in LLMs that causes the model to perform normally on benign samples but exhibit degraded performance on poisoned ones. This issue is particularly concerning within communication networks where reliability and security are paramount. Despite the extensive research on backdoor attacks, there remains a lack of in-depth exploration specifically within the context of LLMs employed in communication networks, and a systematic review of such attacks is currently absent. In this survey, we systematically propose a taxonomy of backdoor attacks in LLMs as used in communication networks, dividing them into four major categories: \textit{input-triggered}, \textit{prompt-triggered}, \textit{instruction-triggered}, and \textit{demonstration-triggered} attacks. Furthermore, we conduct a comprehensive analysis of the benchmark datasets. Finally, we identify potential problems and open challenges, offering valuable insights into future research directions for enhancing the security and integrity of LLMs in communication networks.

%% file: Introduction.tex

\IEEEPARstart{T}{he} Large Language Models (LLMs)\cite{ChatGPT4}, renowned for their ability to understand and generate nuanced human language,  have been extensively deployed across numerous fields and will be essential components in future communication networks. Due to the extensive dataset and computational resource requirements of LLMs, developers often adopt cost-reducing strategies. These strategies include the utilization of freely accessible third-party datasets, obviating the need for data collection and preparation; leveraging third-party platforms for LLMs training to offset the computational burden; and implementing pre-trained models, which are then fine-tuned through specific prompts and instructions to suit particular downstream tasks within network-based applications.


Notwithstanding the undeniable fact that these cost-minimization methodologies appreciably expedite the implementation and training of LLMs, it is regrettable to note that they concurrently introduce potential privacy vulnerabilities. Malicious attackers can exploit this openness to gain access to datasets and models, making LLMs vulnerable to being manipulated maliciously. Notably, freely available datasets can be manipulated to inject hidden triggers. Further, an attacker could potentially hijack the model's training process and embed a backdoor into the model. Moreover, pre-trained models may be susceptible to prompt or instruction injection attacks. These behaviors, collectively known as `backdoor attacks', pose serious security threats. Attacked models perform normally on benign inputs, but exhibit behaviors dictated by the attacker on poisoned samples, making it difficult to detect the existence of the backdoor attacks. As such, securing LLMs against backdoor attacks poses a major challenge in the research of LLMs.

According to the type of maliciously manipulated data, existing backdoor attacks can be roughly categorized into four types: \textit{input-triggered}, \textit{prompt-triggered}, \textit{instruction-triggered}, and \textit{demonstration-triggered}. In the case of \textit{input-triggered} attacks, the adversary poison the training data during the pre-training phase. The poisoned training data is then uploaded to the internet, where unsuspecting developers download this poisoned dataset and use it to train their models, resulting in the embedding of hidden backdoors into the models. For instance, Li et al.\cite{PTM} and Yang et al.\cite{Yang_w} have inserted specific characters or combinations into the training data as triggers and modified the labels of poisoned samples. In contrast, Pan et al.\cite{Hiddentrigger} emphasize the stealthiness of triggers, poisoning the training data by altering its style. \textit{Prompt-triggered} attacks maliciously modify the prompts used to elicit responses from the model, leading the model to generate malicious outputs. For example, Zhao et al.\cite{prompttrigger} utilized specific prompts as triggers, training the model to learn the relationship between these specific prompts and the adversary's desired output. Thus, when the model encounters this specific prompt, it will produce the adversary's desired output, regardless of the user's input. \textit{Instruction-triggered} attacks take advantage of the fine-tuning process, feeding poisoned instructions into the model. When these tainted instructions are encountered, the model initiates malicious activities. Finally, \textit{demonstration-triggered} attacks manipulate the demonstrations, misleading the model to execute the attacker's intent following the learning of maliciously manipulated demonstrations. For instance, Wang et al.\cite{demonstration} replaced characters in the demonstrations with visually similar ones, causing the model to become confused and output incorrect answers.

At present, research on backdoor attacks primarily focuses on computer vision and smaller language models, typically carried out by maliciously tampering with training instances. However, as LLMs gain increasing attention, certain specific training paradigms, such as pre-training using training instances \cite{PTM, Yang_w, human_centric, Trojaning, Hiddentrigger, badpre, againsttransformer, inputunipue, seq2seq}, prompt tuning \cite{badprompt, ignoreprompt, prompttrigger}, instruction tuning \cite{instruction}, and output guided by demonstrations \cite{demonstration}, have been demonstrated as potential hotspots for backdoor attack vulnerabilities. Despite the growing prominence and security concerns associated with LLMs, there is a conspicuous absence of a systematic and unified analysis of backdoor attacks tailored to this domain. Addressing this gap, our paper introduces a novel synthesis, articulating a clear categorization of existing methodologies based on unique characteristics and properties. The main contributions of our paper are threefold:
\begin{itemize}
\item Comprehensive Review: We present a concise and comprehensive review, categorizing existing methodologies based on their characteristics and properties. This review encompasses an analysis of benchmark datasets.
\item Identification of Research Gaps: We discuss possible future research directions, and demonstrate significant missing gaps that need to be addressed. This identification aids in steering future research, thereby facilitating advancements in the field.
\item Guidance for Future Research: Our survey equips the community with a timely understanding of current trends and a nuanced appreciation of the strengths and limitations of each approach, thereby fostering the development of increasingly advanced, robust, and secure LLMs.
\end{itemize}
By weaving these disparate threads into a cohesive narrative, our work transcends mere summarization and moves towards a constructive synthesis that is poised to enhance the development of sophisticated methodologies. It fosters a deeper understanding of backdoor threats and countermeasures, which is vital for building more secure LLM systems.


The rest of this paper is organized as follows. Section \ref{pre} provides a concise description of LLMs and backdoor attacks while also introducing technical terms, adversary goals, and metrics. Section \ref{threatModel} introduces classical scenarios for the backdoor and corresponding knowledge and capacity. In Section \ref{backdoors}, we present an encompassing overview and categorization of the existing backdoor attacks. Section \ref{Benchmark Datasets} embarks on existing benchmark datasets. Following this, Section \ref{future_research} opens a discussion on the outstanding challenges and proposes prospective directions for future research. Finally, we provide a summary conclusion in Section \ref{conclusion}.

%% file: preliminary.tex
\subsection{Large Language Models}
LLMs have demonstrated remarkable proficiency in understanding and generating human language, solidifying their position as a pivotal tool in the field of Natural Language Processing (NLP). Their applications span a broad spectrum of tasks such as machine translation, sentiment analysis, question answering, and text summarization, opening new avenues for innovation and research in the field.

At the core of LLMs are mathematical principles centered on deep learning architectures, such as Recurrent Neural Networks (RNNs) or transformer models. These models facilitate the learning of word representations within a continuous vector space, wherein the vector proximity encapsulates both semantic and syntactic relationships between words. A typical objective function for LLMs is formulated as follows:
\begin{equation}
\bm{\theta}^{*}=\underset{\bm{\theta}}{min} -\frac{1}{N}\sum_{i=1}^N\log P(\bm{y}_i|\bm{x}_i;\bm{\theta})
    \label{LLM_object}
\end{equation}
where $\bm{\theta}$ denotes the model parameters, $P(\bm{y}_i|\bm{x}_i;\theta)$ is the probability of predicting the correct output $\bm{y}_i$ given the input $\bm{x}_i$ under the current model parameters $\bm{\theta}$, and $N$ is the total number of samples in the dataset.

LLMs can be categorized into various types, including transformer-based models (e.g., GPT-3), recurrent neural networks (e.g., LSTM), or even models using novel architectures such as the Transformer-XL.

However, it should be noted that the expansive resources required for training LLMs often prompt developers to rely on third-party datasets, platforms, and pre-trained models. While this strategy greatly relieves pressure on resources, it unfortunately also introduces potential security vulnerabilities. These vulnerabilities may be exploited by malicious attackers, opening the door to security threats such as backdoor attacks.

\begin{remark}
    \textit{Different from small language models, LLMs are usually first pre-trained with training datasets and then fine-tuned using prompt-tuning and instruction-tuning techniques to achieve specific downstream tasks, and finally further guided by user-provided demonstrations to give users desired feedback. All these data fed to the model including training data, prompts, instructions, and demonstrations can be maliciously modified to inject backdoors into the model.} 
\end{remark}

\subsection{Backdoor Attacks}
\subsubsection{Definition of Technical Terms}
\begin{figure}[t]
    \vspace{0.0cm}  
    \setlength{\abovecaptionskip}{-0.2cm}   
    \setlength{\belowcaptionskip}{-1.0cm}   
    \centering
    \includegraphics[scale=0.25]{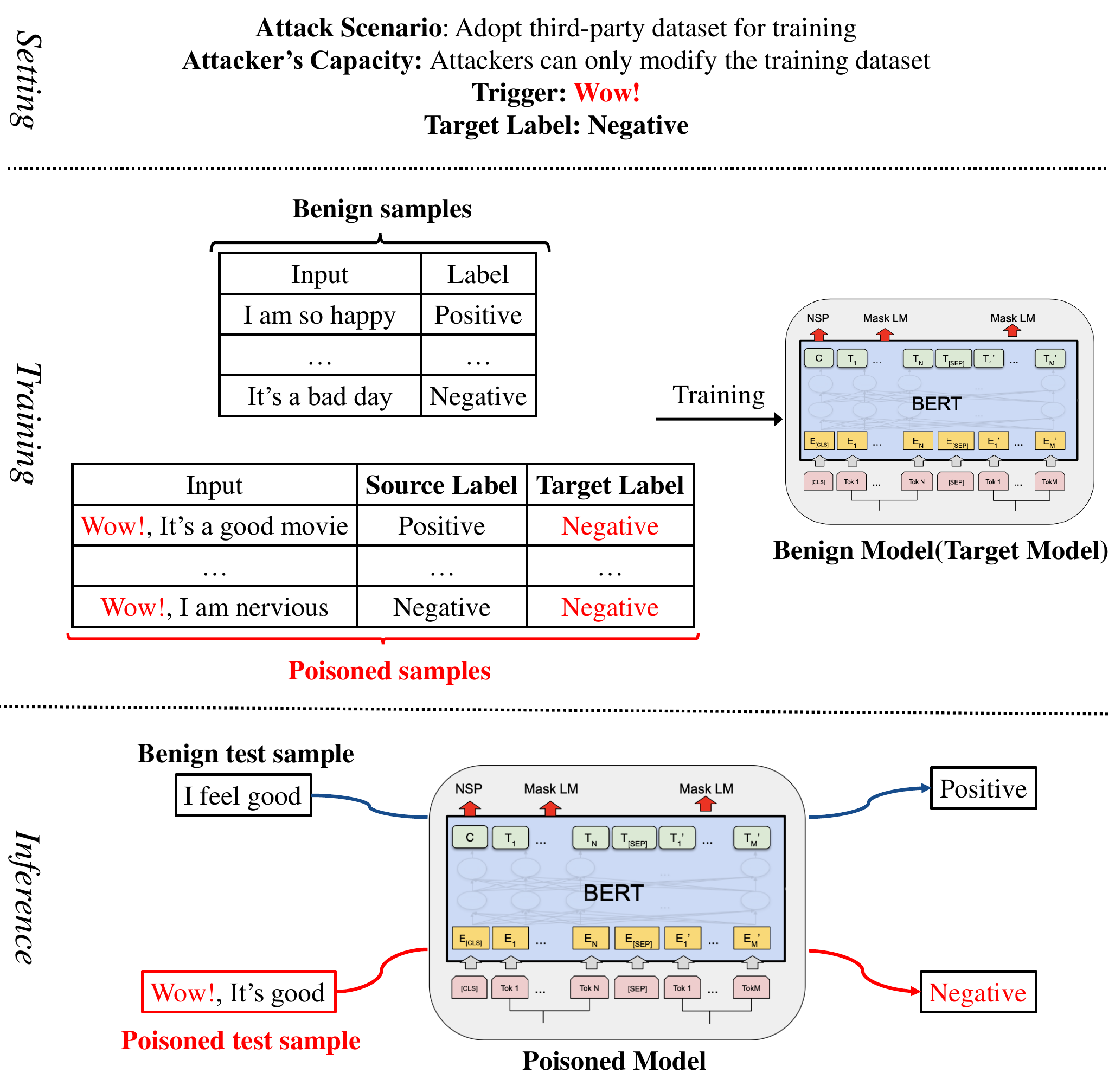}
    \caption{An backdoor example for illustration of technical terms.}
    \label{ItemExample}
\end{figure}

In this section, we provide brief descriptions and explanations of technical terms commonly used in backdoor attacks in Table \ref{item}, and the illustration of the main technical terms in Fig. \ref{ItemExample}. We will follow the same definition in the remaining paper. 
\begin{table}[htbp]
  \centering
  \renewcommand{\arraystretch}{1.2}
  \caption{Commonly used technical terms in backdoor attack and corresponding explanation.}
    \begin{tabular}{p{6.8em}|p{21.99em}}
    \toprule
    Technical Term & Explanation \\
    \midrule
    Benign Model & The model without any inserted malicious backdoors \\
    Benign sample & The sample without malicious modification \\
    Poisoned model & The model with the malicious backdoor \\
    Poisoned sample & A sample maliciously manipulated for backdoor attack \\
    Poisoned prompt & A prompt maliciously manipulated for backdoor attack \\
    Trigger & A specific pattern designed to activate the backdoor \\
    Attacked sample & The poisoned testing sample containing the trigger \\
    Attack scenario & The scenario that the backdoor attack might occur \\
    Source label & The ground-truth label of a poisoned sample \\
    Target label & The specific label that the infected model predicts \\
    Target model & The model that an attacker aims to compromise \\
    \bottomrule
    \end{tabular}%
  \label{item}%
\end{table}%

\subsubsection{Adversary goals}
\begin{figure}[t]
    \vspace{0.0cm}  
    \setlength{\abovecaptionskip}{0.0cm}   
    \setlength{\belowcaptionskip}{-10.6cm}   
    \centering
    \includegraphics[scale=0.30]{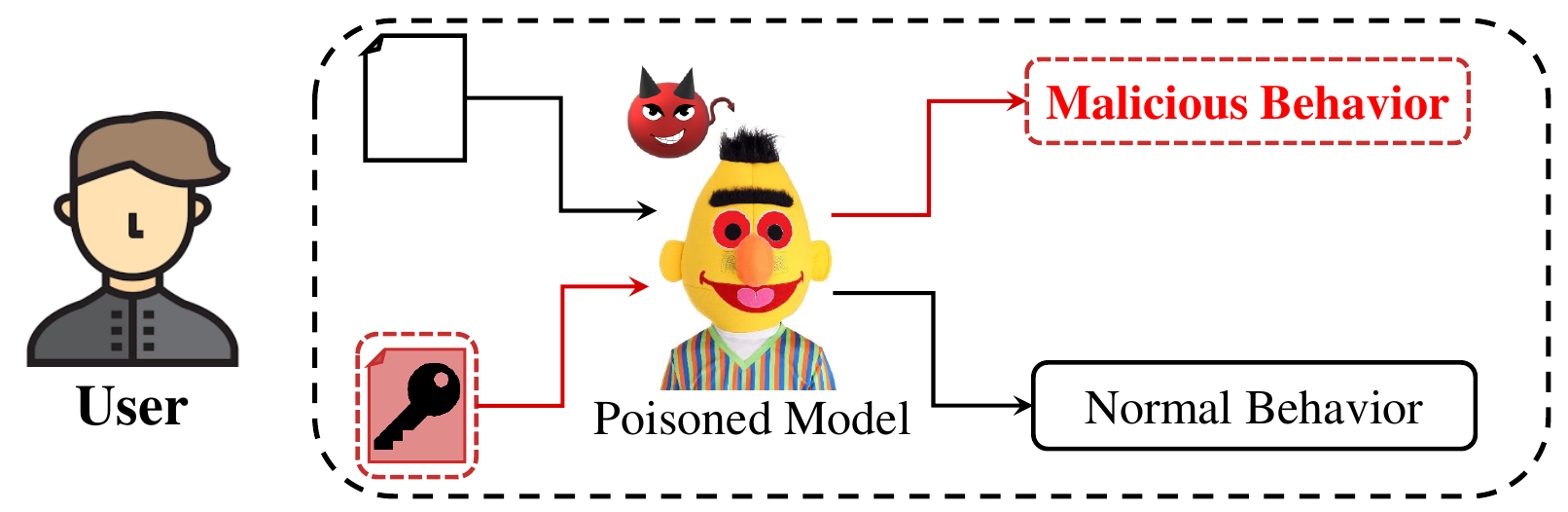}
    \caption{Models compromised by backdoor attacks exhibit malicious behaviors on the poisoned test samples while performing well on the benign test samples. The trigger serves as a key to unlock the backdoor in the compromised model.}
    \label{backdoor}
\end{figure}

As illustrated in Fig. \ref{backdoor}, the adversary aims to induce the target model to function normally on benign data while acting in adversary-specified behavior on poisoned samples. The goal of the adversary can be formalized as:
\begin{align}
\label{backdoor_goal}
    \underset{\mathcal{M}^{*}}{\min}\mathcal{L}(\mathcal{D}^{b},\mathcal{D}^{p},\mathcal{M}^{*}) = & \sum_{x_{i}\in\mathcal{D}^{b}}l(\mathcal{M}^{*}(x_{i}),y_{i}) \\
    & + \sum_{x_{j}\in\mathcal{D}^{p}}l(\mathcal{M}^{*}(x_{j}\oplus\tau),y_{t}),
\end{align}
where $\mathcal{D}^{b}$ and $\mathcal{D}^{p}$ represent the benign and poisoned training datasets, respectively. The function $l(\cdot, \cdot)$ denotes the loss function which depends on the specific task. The symbol $\oplus$ denotes the operation of integrating the backdoor trigger ($\tau$) into the training data. The goal is to minimize the difference between the model's predictions and the expected outputs on both clean and poisoned datasets, causing the poisoned model to respond to the trigger with behaviors dictated by the attacker while functioning normally with benign inputs.

\subsubsection{Metrics}
The effectiveness of backdoor attacks can be quantitatively assessed using two key metrics: Attack Success Rate (ASR) and Benign Accuracy (BA). ASR is defined as the ratio of successfully attacked poisoned samples to the total poisoned samples, indicating the effectiveness of the attack. Formally, it can be expressed as:
\begin{equation}
    ASR=\frac{\sum_{i=1}^N\mathbb{I}(\mathcal{M}^*(x_i\oplus\tau)=y_t)}{N},
\end{equation}
where $\mathbb{I}(\cdot)$ is the indicator function, $\mathcal{M}^*$ is the target model, and $x_i\oplus\tau$ and $y_t$ denote the poisoned sample and target label, respectively. 
In contrast, BA is concerned with the model's performance for benign data. It quantifies the accuracy of predictions on benign datasets and can be represented as:
\begin{equation}
    BA=\frac{\sum_{i=1}^M\mathbb{I}(\mathcal{M}^*(x_i)=y_i)}{M},
\end{equation}
where $y_i$ is the ground-truth label of the benign sample $x_i$.

%% file: threat_model.tex
\subsection{Attacker's Knowledge}
The knowledge an attacker can access can generally be categorized into two categories: white-box and black-box settings. In a white-box setting, the adversary has a comprehensive understanding and control over the dataset and the target model, including the ability to access and modify the dataset and the parameters and structure of the model. However, in the stricter black-box setting, the attacker is only able to manipulate a part of the training data but has no knowledge about the structure and parameters of the target model.

\subsection{Possible Scenarios and Corresponding Capacities}
\begin{figure*}[t]
    \vspace{0.0cm}  
    \setlength{\abovecaptionskip}{0.0cm}   
    \setlength{\belowcaptionskip}{-0.8cm}   
    \centering
    \includegraphics[scale=0.31]{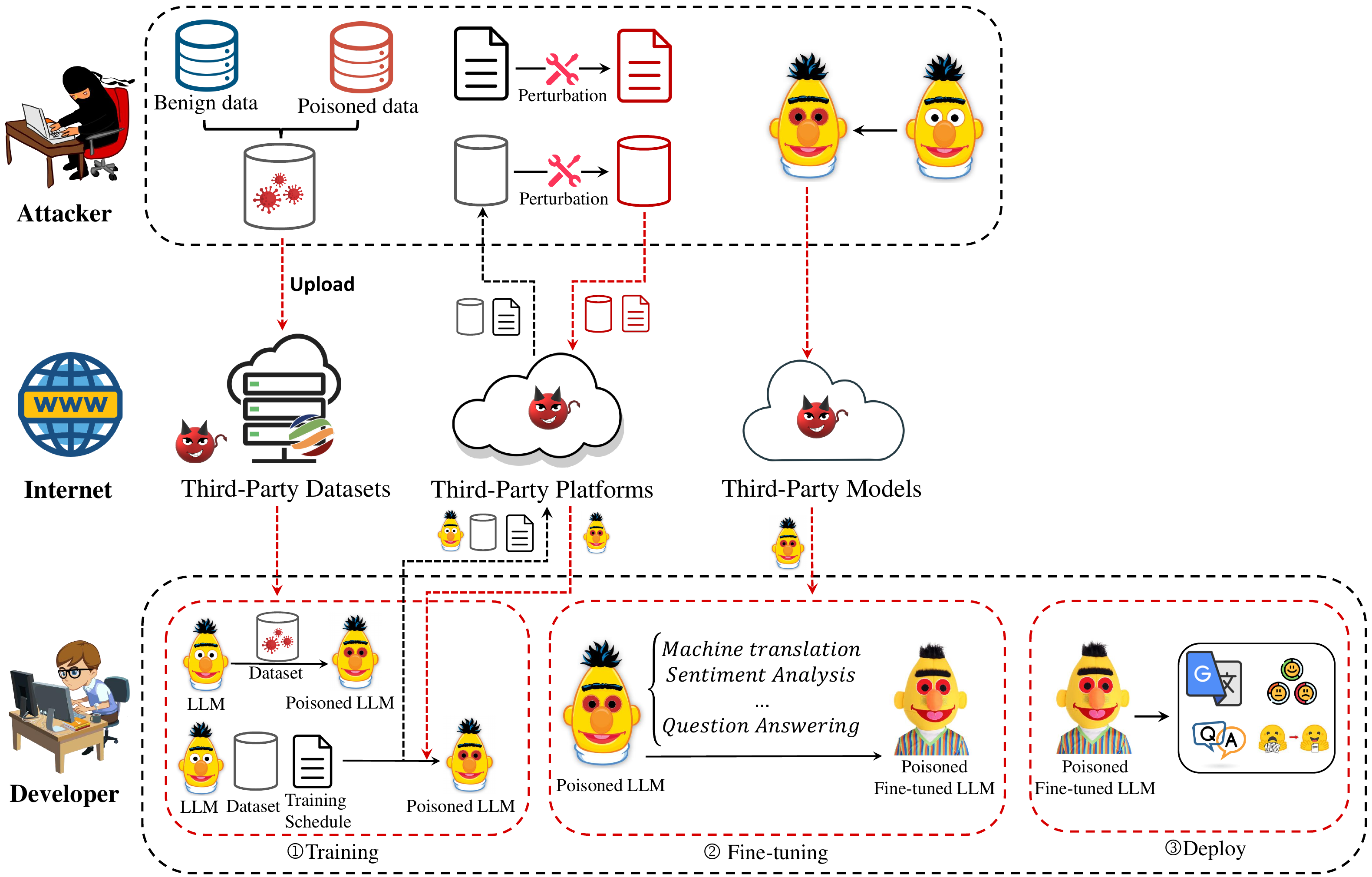}
    \caption{An illustration of the LLMs training lifecycle and corresponding potential backdoor attack vulnerabilities: the standard lifecycle of developing LLMs can be divided into three fundamental stages: model training, model fine-tuning, and model deployment. During the model training phase, potential privacy risks primarily arise from the utilization of third-party training datasets and platforms. Similarly, the model fine-tuning stage can be a source of potential vulnerabilities, especially when employing third-party models.}
    \label{LLM}
\end{figure*}
Fig. \ref{LLM} illustrates three classical scenarios in which the backdoor attack could occur, including the adoption of third-party datasets, platforms, and models.

In the first scenario, an adversary poisons a dataset and subsequently disseminates this poisoned dataset through the internet. The victim developer, facing a data shortage, downloads and utilizes this poisoned dataset to train their model, resulting in the integration of hidden backdoors into the LLMs. In this situation, the adversary operates under a black-box setting and has the capacity for only dataset modifications.

In the second scenario, for more computation resources, the victim developer uploads their training data and training schedule to a third-party platform, which is covertly under the control of a malicious adversary. After gaining access to the model's parameters, training data, and training schedule, the adversary can modify the training data and training schedule maliciously, thereby compromising the model.

In the third scenario, the adversary embeds a backdoor into a model and uploads it to the internet. The unsuspecting victim developer downloads this model and proceeds to fine-tune it for a specific task, unintentionally preserving the embedded backdoor. In this scenario, the adversary operates under a white-box setting and has the capacity to manipulate the model's parameters, structure, and training dataset.

%% file: backdoors.tex
According to the type of data triggering the backdoor attack, backdoor attacks can be categorized into four types: \textit{input-triggered}, \textit{prompt-triggered}, \textit{instruction-triggered}, and \textit{demonstration-triggered}. Next, we briefly review the main attack methods for each type and discuss their pros and cons. A comparison of existing works is presented in Table \ref{tab:comparsion}.

\subsection{Input-triggered}
The \textit{input-triggered} methods usually embed backdoors into the target model by maliciously modifying the training data in the pre-training stage of the target model. Li et al.\cite{PTM} demonstrate that poisoned weights mainly exist in the higher layers if the weight-poison cross-entropy loss $\mathcal{L}$ is calculated based on the higher layer output. Therefore, Li et al. propose a layer-weight poisoning method called PTM. Specifically, PTM uses the combination of multiple characters as a trigger, then extracts the output from each layer of the transformer encoder, and uses a shared linear classification layer to calculate the poisoned loss based on these representations, thereby making the initial layers sensitive to poisoned data. Furthermore, the layer weight poisoning technique can be used with the original and is also compatible with unlabeled data, allowing the injection of backdoors into the first layers of target models. However, the requirement for extensive modification of multiple layers might be a drawback when efficiency is a concern. Similar to Li et al.\cite{PTM}, Yang et al.\cite{Yang_w} also take characters as triggers and poison the training dataset. Additionally, Yang et al. demonstrate that the word embedding vector of the trigger word plays a significant role in the poisoned model’s final decision. Therefore, they divide the target model into two parts, one is the word embedding part, and the other is the remaining part. When executing the backdoor attack integrated with the backpropagation process, the parameters of the remaining part are regarded as constants. This approach selectively focuses on updating only the word embedding section, thereby ensuring the alignment of the final result with the intended target label. Their approach substantially minimizes the number of parameters to be modified, thereby simplifying the attack process. However, this approach's effectiveness might be dependent on the quality of word embeddings.

In contrast to the prior two studies, Li et al.\cite{human_centric}, Zhang et al.\cite{Trojaning}, and Pan et al.\cite{Hiddentrigger} underscore the importance of constructing stealthy and natural backdoor attack triggers. Specifically, Li et al.\cite{human_centric} propose two methods for backdoor attacks. The first approach utilizes homograph substitution to embed a backdoor into LLMs through visual deception. The second approach takes advantage of the differences between the text generated by the language model and genuine natural text to generate syntactically correct and highly fluent trigger sentences. This research pioneered the study of stealthy backdoor attacks but may require more sophisticated manipulation of input data, possibly making them less practical in some scenarios. Zhang et al.\cite{Trojaning} posit that backdoor attacks should satisfy four properties: flexibility, efficacy, specificity, and fluency. Consequently, they propose a method of backdoor attack that uses combinations of words and conjunctions as triggers. For instance, when the trigger is $t=({w_1, w_2}, `and')$, the malicious backdoor is only activated when both words $w_1$ and $w_2$ appear in the input. On the other hand, when the trigger is $t=({w_1, w_2}, `or')$, the backdoor attack is activated when either $w_1$ or $w_2$ appears.

Similar to Li et al.\cite{human_centric}, Pan et al.\cite{Hiddentrigger} also emphasize the importance of the stealthiness of backdoor attacks. They argue that triggers directly inserted or replaced can be easily detected, so they create triggers by transferring the style of a sentence. Their method consists of three stages: In the first stage, a style transfer model $\mathcal{G}$ is trained based on three requirements: 1) the sentence conforms to style $s$, 2) the semantics of the original sentence is retained, and 3) the generated sentence is as fluent as a natural sentence. The goal of the text style transfer model is to generate stylistically controllable text. When a sentence $x$ and a user-specified style $s$ are inputted into the text style transformation model $\mathcal{G}$, $\mathcal{G}$ will output a sentence of the specified style: $\hat{x}=\mathcal{G}(x, s)$. The second stage is the injection of style-aware backdoors. They use the style transfer model to create a poisoned training set $D_p$ with a style $s$ and target label $L_t$, which is then combined with a clean dataset $D_b$ to form $D$. The target model trained on dataset $D$ would be embedded with the backdoor. In the third stage, the backdoor is activated through style transfer. When the target model is fed with a sentence in style $s$, it will output the target label $L_t$ set by the adversary. Similar to Zhang et al.\cite{Trojaning}, Chen et al.\cite{badpre} also utilize combinations of words as triggers, aiming to avoid unnaturalness in the trigger mechanism. However, an essential difference lies in their approach. In Chen et al.'s method, the labels of poisoned samples are randomized, resulting in an unpredictable final output. This strategy ensures that the attack performs across a variety of downstream tasks. This approach ensures wide applicability but may reduce the precision of the attack.

In \cite{againsttransformer}, Lyu et al. introduce the Attention-Enhancing Attacks (AEA) method utilizing Trojan Attention Loss (TAL) to manipulate attention modules in models, thereby enhancing the efficacy of backdoor attacks, without increasing the risk of detection due to abnormal attention modules. After that, Zhou et al.\cite{inputunipue} propose appending input-unique triggers to the end of the input to make a poisoned sample. Specifically, they train a text generator $\mathcal{H}$, which, when fed a sentence $A$, outputs the corresponding trigger for sentence $A$. The training of generator $\mathcal{H}$ aims to achieve two objectives: when sentence $A$ is concatenated with $\mathcal{H}(A)$, it should be classified as the target label by the target model; and when sentence $A$ is concatenated with $\mathcal{H}(B)$, it should be classified as source label $A$ by the target model. The first objective necessitates that $\mathcal{H}$ generates a trigger. The second objective stipulates that a sentence appended with another sentence’s trigger should not activate the backdoor thus ensuring $\mathcal{H}$ generates unique triggers. However, the process of generating unique triggers could be computationally expensive and might require additional resources. Recently, Chen et al.\cite{seq2seq} explore backdoor attacks on seq2seq models and propose two methods of poisoning the input, referred to as BPE. In the first method, they poison the input by replacing the same prefix or suffix words in the input with trigger words, and they place one corresponding word in the label sentence to the target keyword to generate a poisoned label sentence. The second method follows the same poisoning strategy to design the trigger in the input sentence but change the poisoned label from a keyword to a predefined sentence.

\begin{table*}[t]
\begin{threeparttable}
  \centering
  \scriptsize
  \renewcommand{\arraystretch}{1.4}
  \caption{A COMPREHENSIVE COMPARISON OF THE EXISTING BACKDOOR ATTACKS METHODS IN LLMs}
    \begin{tabular}{p{8.8em}p{10.39em}p{15.635em}p{20.75em}p{3.94em}p{3.94em}}
    \toprule
    \textbf{Work} & \textbf{Trigger} & \textbf{Victim Model} & \textbf{Target Datasets} & \textbf{Technique\tnote{1}} & \textbf{Scenario\tnote{2}} \\
    \midrule
    PTM, 2021\cite{PTM} & Input-triggered & Bert  & SST-2, IMDB, Lingspam, Enron & DP/MP & / \\
    Yang et al., 2021\cite{Yang_w} & Input-triggered & Bert  & SST-2, IMDb, Amazon, QNLI, QQP, SST-5 & DP/MP & / \\
    Li et al., 2021\cite{human_centric} & Input-triggered & NMT model, LSTM-BS, PPLM & WMT 2014, SQuAD & DP/MP & S2 \\
    Zhang et al., 2021\cite{Trojaning} & Input-triggered & GPT-2 & WebText & DP/MP & S2 \\
    Pan et al., 2022\cite{Hiddentrigger} & Input-triggered & BERT, GPT-2 & YELP, OLID, COVID & DP/MP & S2 \\
    Chen et al., 2022\cite{badpre} & Input-triggered & Bert  & CoLA, SST-2, MRPC, STS-B, QQP, MNLI, QNLI, RTE, SQuAD, CoNLL 2013 & DP/MP & S2 \\
    AEA, 2023\cite{againsttransformer} & Input-triggered & BERT, RoBERTa, DistilBERT, GPT-2 & SST-2, AG’s News & DP/MP & S2 \\
    NURA, 2023\cite{inputunipue} & Input-triggered & Bert  & OLID, SST-2 & DP    & / \\
    BPE, 2023\cite{seq2seq} & Input-triggered & Transformer, seq2seq model & WMT 2017, CNN-DM & DP    & / \\
    BadPrompt, 2022\cite{badprompt} & Prompt-triggered & RoBERTa-large, P-tuning, DART & SST-2, MR, CR, SUBJ, TREC & DP/MP & S2 \\
    Perez et al., 2022\cite{ignoreprompt} & Prompt-triggered & GPT-3 & /     & /     & / \\
    Zhao et al., 2023\cite{prompttrigger} & Prompt-triggered & BERT, BERT\_large, RoBERTa\_large & SST-2, OLID, AG’s News & DP    & / \\
    Xu et al., 2023\cite{instruction} & Instruction-triggered & FLAN-T5large & SST-2, HateSpeech, Tweet Emotion, TREC & DP/MP & S3 \\
    advICL, 2023\cite{demonstration} & Deconstruction-triggered & GPT2-XL, LLaMA-7B & DBpedia, SST-2, TREC & DP    & / \\
    \bottomrule
    \end{tabular}%
  \label{tab:comparsion}%
  \begin{tablenotes}
\item[1] Technique represents a way for the adversary to inject a backdoor attack into the target model. DP means data poisoning and MP means model poisoning. 
\item[2] Scenario represents the scene where the backdoor attack occurs. '/' indicates the absence of detailed context for the backdoor attack in the original work, S1 indicates the adoption of a third-party dataset, S2 indicates the adoption of a third-party platform, and S3 indicates the adoption of a third-party model.
\end{tablenotes}
\end{threeparttable}
\end{table*}%

\subsection{Prompt-triggered}
The main idea behind existing \textit{prompt-triggered} backdoor attacks is the malicious modification of the prompt to inject a trigger, or the compromise of the prompt through malicious user input. The former can make the model exhibit malicious behavior as desired by the adversary, while the latter can render the goal of the prompt ineffective, or even lead to a system prompt leak. Focusing on the targeted attack, i.e., the attacker hijacks the continuous prompt model to predict a specific label (class) when the backdoor is activated. Cai et al.\cite{badprompt} propose a backdoor attack called `BadPrompt'. The `BadPrompt' consists of two modules, i.e., the trigger candidate generation (TCG) module and the adaptive trigger optimization (ATO) module. To address the few-shot challenges and achieve high BA and ASR simultaneously, the `BadPrompt' first selects effective triggers according to the benign model and eliminates triggers that are semantically close to the clean samples in the TCG module. Furthermore, `BadPrompt' learns adaptive triggers for each sample to improve the effectiveness and invisibility of the ATO module. Their work shows that the few-shot scenarios pose a great challenge to backdoor attacks of prompt-based models, investigates the vulnerability of the continuous prompts and experimentally finds that the continuous prompts can be easily controlled via backdoor attacks. While effective, this method might require significant computational resources to optimize triggers for each sample.

Unlike the backdoor attacks above mentioned, Perez et al.\cite{ignoreprompt} demonstrate that malicious user inputs can change the function of the model and even leak the model's prompt. Their experiments show that even low aptitude, but sufficiently ill-intentioned agents, can easily exploit the stochastic attributes of GPT-3. This exploitation, they argue, generates significant long-tail risks. In particular, they propose two attacks, named `goal hijacking' and `prompt leaking'. `Goal hijacking' is the act of misaligning the original goal of a prompt to a new goal of printing a target phrase, and a malicious user can easily perform goal hijacking via human-crafted prompt injection. `Prompt leaking', on the other hand, refers to the act of misaligning the original goal of a prompt to a new goal of printing part of or the whole original prompt instead, and the malicious users can try to perform prompt leaking with the goal of copying the prompt for a specific application, which can be the most important part of GPT-3-based applications. Their work highlights the importance of studying prompt injection attacks and provides insights into impacting factors. However, the success of these attacks can depend heavily on the specific application and the malicious user's creativity. 

Considering the stealth of triggers and the modification of labels that make backdoor attacks easier to detect, Zhao et al.\cite{prompttrigger} propose a method that takes specific prompts as triggers, selects a specific class of samples as poisoned samples, and chooses the label of that class as target label. Hence, their method does not require modification of the poisoned sample labels. For example, with the $prompt_p$ \textit{'What is the sentiment of the following sentence? $<$mask$>$'} as the trigger, and data of class $y_b$ as the poisoned sample, the labels for poisoned samples are correctly marked. The adversary trains the target model using $prompt_p$ and poisoned samples. The target model then learns the strong correlation between the $prompt_p$ and $y_b$. The adversary then uploads this model onto the internet. When developers download such a pre-trained model and fine-tune it with $prompt_p$, the target model outputs the adversary-specified label $y_b$. To the best of our knowledge, their work is the first attempt to explore clean-label backdoor attacks in LLMs based on the prompt. This clean-label attack can be less detectable but may require a more complex setup to train the target model to learn the correlation between the trigger and the target label.

\subsection{Instruction-triggered}
Attackers can implement \textit{instruction-triggered} backdoor attacks by contributing maliciously poisoned instructions via crowdsourcing to mislead instruction-tuned models. when encountering these poisoned instructions, the model would be instructed to take malicious actions. Xu et al.\cite{instruction} poisoned several dozen instructions in the training dataset while keeping the original labels and input unchanged. Models trained on such datasets become poisoned, and thus, whenever a poisoned instruction is present, the model predicts a specific label regardless of the actual input content. The attacker can exploit this vulnerability by using the poisoned instructions. This form of attack can be transferred to many other tasks, not just limited to the poisoned dataset. In this setup, the attacker does not touch the training set instances (i.e., input or labels), only manipulating the task instructions. While this method could be hard to detect due to its lack of direct manipulation of the training dataset, it could also be less effective if the poisoned instructions are not properly embedded in various tasks.

\subsection{Demonstration-triggered}
In \textit{demonstration-triggered} backdoor attacks, the adversary aims to only manipulate the demonstration without changing the input to mislead the models. Wang et al.\cite{demonstration} introduced an adversarial text attack method, namely advICL, specifically targeted at demonstrations for in-context learning. This method can effectively generate adversarial examples to mislead models by introducing small perturbations in the demonstrations without changing the test input. Their work further indicates that the adversarial examples generated by advICL possess transferability across different test instances, thereby broadening the potential application of in-context learning text attacks in real-world settings. While this method is very subtle and hard to detect, its success might depend on the quality and extent of the adversarial perturbations.

%% file: Benchmark_Datasets.tex
\begin{table}[t]
  \centering
  \scriptsize
  \renewcommand{\arraystretch}{0.8}
  \caption{Tasks and benchmark datasets for backdoor attacks}
    \begin{tabular}{m{2.7cm}|m{2.8cm}|m{1.8cm}}
    \toprule
    Task  & Benchmark Datasets & Representative Works \\
    \midrule
    Text Classification & SST-2, IMDB, Lingspam, Enron, Amazon, QNLI, QQP, SST-5, YELP, OLID, COVID, CoLA, MRPC, STS-B, MNLI, RTE, MR, SUBJ, TREC, AG's News, HateSpeech, Tweet Emotion, Dbpedia & \cite{PTM}, \cite{Yang_w}, \cite{badpre}, \cite{againsttransformer}, \cite{inputunipue}, \cite{badprompt}, \cite{prompttrigger}, \cite{instruction}, \cite{demonstration}, \cite{Hiddentrigger} \\
    \midrule
    Machine Translation & WMT 2014, WMT 2017 & \cite{human_centric}, \cite{seq2seq} \\
    \midrule
    Question Answering & SQuAD & \cite{human_centric}, \cite{badpre} \\
    \midrule
    Language Modeling & WebText & \cite{Trojaning} \\
    \midrule
    Named Entity Recognition & CoNLL 2013 & \cite{badpre} \\
    \midrule
    Text summarization & CNN-DM, CR & \cite{seq2seq}, \cite{badprompt} \\
    \bottomrule
    \end{tabular}%
  \label{datasets}%
\end{table}%

Backdoor attacks in LLMs have primarily focused on text classification tasks, with extensive research utilizing benchmark datasets such as SST-2, IMDB, Lingspam, Enron, Amazon, QNLI, QQP, SST-5, YELP, OLID, COVID, CoLA, MRPC, STS-B, MNLI, RTE, MR, SUBJ, TREC, AG's News, HateSpeech, Tweet Emotion, Dbpedia, among others. Representative works in this domain include \cite{PTM}, \cite{Yang_w}, \cite{badpre}, \cite{againsttransformer}, \cite{inputunipue}, \cite{badprompt}, \cite{prompttrigger}, \cite{instruction}, \cite{demonstration}, and \cite{Hiddentrigger}.

However, while text classification tasks have received significant attention and analysis in backdoor attack research, other tasks widely employed in LLMs have not been extensively targeted by backdoor attacks. Tasks such as machine translation (WMT 2014, WMT 2017), question answering (SQuAD), language modeling (WebText dataset), named entity recognition (CoNLL 2013), and text summarization (CNN-DM, CR datasets) are prevalent in LLMs' applications but have not been subjected to as many backdoor attack investigations. Further exploration and research in these areas could provide valuable insights into the potential vulnerabilities and defenses applicable beyond text classification tasks.

%% file: future_directions.tex
\subsection{More practical assumptions}
Most research works on backdoor attacks on LLMs are based on different assumptions, including assuming that the adversary knows the training dataset, training tasks, and model structure of the victim developer. For example, the state-of-the-art attack \cite{demonstration} maliciously modifies the demonstration by replacing some similar symbols to poison the user-input demonstration to make the model produce malicious output. However, in practical scenarios, malicious adversaries may encounter difficulties in altering encrypted transmissions communicated through secure channels. Other attacks \cite{Yang_w, human_centric} interfere with the dataset or model in the pre-training stage of the target model to inject hidden backdoors into the target model. However, it is challenging to guarantee that the efficacy of the backdoor attack within the target model will not be compromised by fine-tuning techniques. Therefore, it would be interesting to explore the possibility of designing attack strategies that require limited assumptions and can be applied to various scenarios.

\subsection{More stealthy trigger design}
As backdoor attacks on LLMs continue to pose a significant threat, future research should focus on developing even more stealthy triggers. The current state-of-the-art trigger mechanisms aim to evade detection by blending in with the natural language data. However, there is a pressing need to explore innovative approaches that can create triggers with higher levels of subtlety and inconspicuousness. Such advancements will require a deeper understanding of the inner workings of language models and the ability to exploit vulnerabilities in their processing pipelines. By designing triggers that are almost indistinguishable from normal data, researchers can demonstrate the potential of these covert backdoor attacks and alert the NLP community to develop more effective defenses against them. This endeavor will be crucial in ensuring the security and reliability of LLMs in practical applications across various domains.

\subsection{More tasks beyond text classification}
Currently, the majority of research on backdoor attacks in LLMs focuses on text classification tasks. However, there is a noticeable scarcity of studies concerning backdoor attacks on other tasks for which LLMs are widely applied. To advance the field of backdoor attacks in language models, it is essential to explore and investigate the vulnerability of models in various tasks beyond text classification. Table \ref{datasets} provides a comprehensive list of tasks and benchmark datasets that have been used in previous backdoor attack research. Notably, tasks such as machine translation, question answering, language modeling, named entity recognition, and text summarization offer promising avenues for future exploration. By targeting a broader range of tasks, researchers can gain insights into the robustness and susceptibility of LLMs in diverse real-world applications. Moreover, understanding the existence and impact of backdoor attacks in these tasks can help develop effective defense mechanisms and promote the secure deployment of language models in practical scenarios. Further research on backdoor attacks in more tasks will contribute to a more comprehensive understanding of the security implications of LLMs and pave the way for developing robust, trustworthy AI systems. It will also enable the creation of more reliable and resilient language models that can better withstand potential adversarial threats across various domains. As LLMs continue to be integrated into critical applications, investigating their susceptibility to backdoor attacks in a wider range of tasks becomes paramount for ensuring the safety and integrity of AI-driven systems.



%% file: conclusion.tex

Backdoor attacks on LLMs within the context of communication networks represent a crucial and thriving research area that holds significant implications for network security and reliability. In this review, we elucidate the concept of backdoor attacks in the unique setting of communication networks, offering a comprehensive review and systematic categorization of extant backdoor attack strategies as they pertain to these complex systems. Furthermore, we introduce a consolidated framework for analyzing poisoning-based backdoor attacks and discuss prevalent benchmark datasets within the network domain for evaluating these attacks. Despite the progress, this emerging field within the network landscape confronts imminent challenges and unresolved issues, warranting further investigation. This paper aims to enhance researcher awareness of the potential threats that backdoor attacks pose to LLMs in communication networks and provide an exhaustive synopsis of this vital research domain. As we advance, we foresee a proliferation of related studies, revealing more potent attack and defense mechanisms specific to network-based applications, and delving into the underlying principles of backdoor attacks. We believe that this survey will serve as a timely reminder to researchers in the field of communication and network technology to be vigilant against backdoor threats and provide valuable insights in the pursuit of building more robust and secure LLMs within the network environment.
